\documentclass[twocolumn,pra,aps,floats,showpacs,amsmath,amssymb]{revtex4}
\usepackage{graphicx}
\usepackage{bm}

\begin{document}
\newcommand{\re}{\,{\rm Re}\,}
\newcommand{\im}{\,{\rm Im}\,}
\newcommand{\tr}{\,{\rm tr}\,}

\title{Excess quantum noise due to mode nonorthogonality in dielectric microresonators}

\author{Henning Schomerus}
\affiliation{Department of Physics, Lancaster University, Lancaster, LA1 4YB,
United Kingdom}

\date{June 2009}
\begin{abstract}
This work presents a theory of the frequency-resolved light emission of
active two-dimensional dielectric microresonators, which are  characterized
by a highly non-paraxial mode structure and frequently feature  a
position-dependent dielectric constant and non-uniform gain. The Lorentzian
intensity profile is characterized by an appropriately generalized Petermann
factor, a renormalized peak position, and the cold-cavity resonance lifetime
$\Gamma$. The theory also delivers a relation of $\Gamma$ to the laser
threshold that improves earlier phenomenological expressions even for the
case of a homogeneous medium.
\end{abstract}

 \maketitle

Breit-Wigner theory is a cornerstone of the study of open quantum
systems and finds numerous applications from atomic and nuclear
physics to mesoscopic transport and optical or acoustical
resonators. This perturbative theory starts from the isolated
situation and delivers a relation between the total scattering
cross section and the resonance linewidth $\Delta\omega$. Because
of its perturbative nature, Breit-Wigner theory breaks down for
moderately open systems. A context which is particularly suited to
explore the ensuing new physics are lasers, which are known to
emit light around a sharply defined frequency $\bar\omega$ with a
Lorentzian intensity profile,
\begin{equation} I(\omega)=\frac{1}{2\pi}\frac {I
\Delta\omega}{(\omega-\bar\omega)^2+\Delta\omega^2/4}.
\label{eq:lorentzian}
\end{equation}
Here $I=\int d\omega\,I(\omega)$ is the total output intensity.
The linewidth is ultimately limited by the spontaneous emission of
photons. For almost lossless resonators, the Schawlow-Townes
formula relates the quantum-limited linewidth $\Delta\omega_{\rm
ST}=\Gamma^2/2I$ to the total output intensity $I$ and the
cold-cavity decay rate $\Gamma$ \cite{st}. This prediction is
based on Breit-Wigner theory, and amounts to one spontaneously
emitted noise photon per cavity mode. The explicit factor of $1/2$
accounts for the amplitude suppression of field fluctuations due
to the active feedback of the medium, and is absent below the
lasing threshold.

That the sensitivity to quantum noise increases for more open
resonators was first realized by Petermann  \cite{petermann}, who
considered gain-guided semiconductor lasers. He predicted that the
linewidth is enhanced by a factor of $K\geq 1$ such that
$\Delta\omega=K\Gamma^2/2I$. Siegman \cite{siegman} developed a
general framework to study open-sided resonators with a paraxial
mode structure and arrived at the expression
\begin{equation} K_{\rm 0}=\left|\frac{\int |\psi_0|^2 d{\bf r} }{ \int \psi_0^2 d{\bf
r}}\right|^2\label{eq:ksp}
\end{equation}
 which relates the
Petermann factor to the inverse condition number of the transverse
resonance wavefunction $\psi_0$. This establishes a direct
connection to a mathematical measure of mode nonorthogonality.
Subsequently, several groups demonstrated the excess noise in
experiments on various open-sided resonator geometries with
different cross-section
\cite{experiment1,experiment2,experiment3,experiment4,experiment5}.

\begin{figure}
\hspace*{-3.27cm}\includegraphics[width=2.472cm]{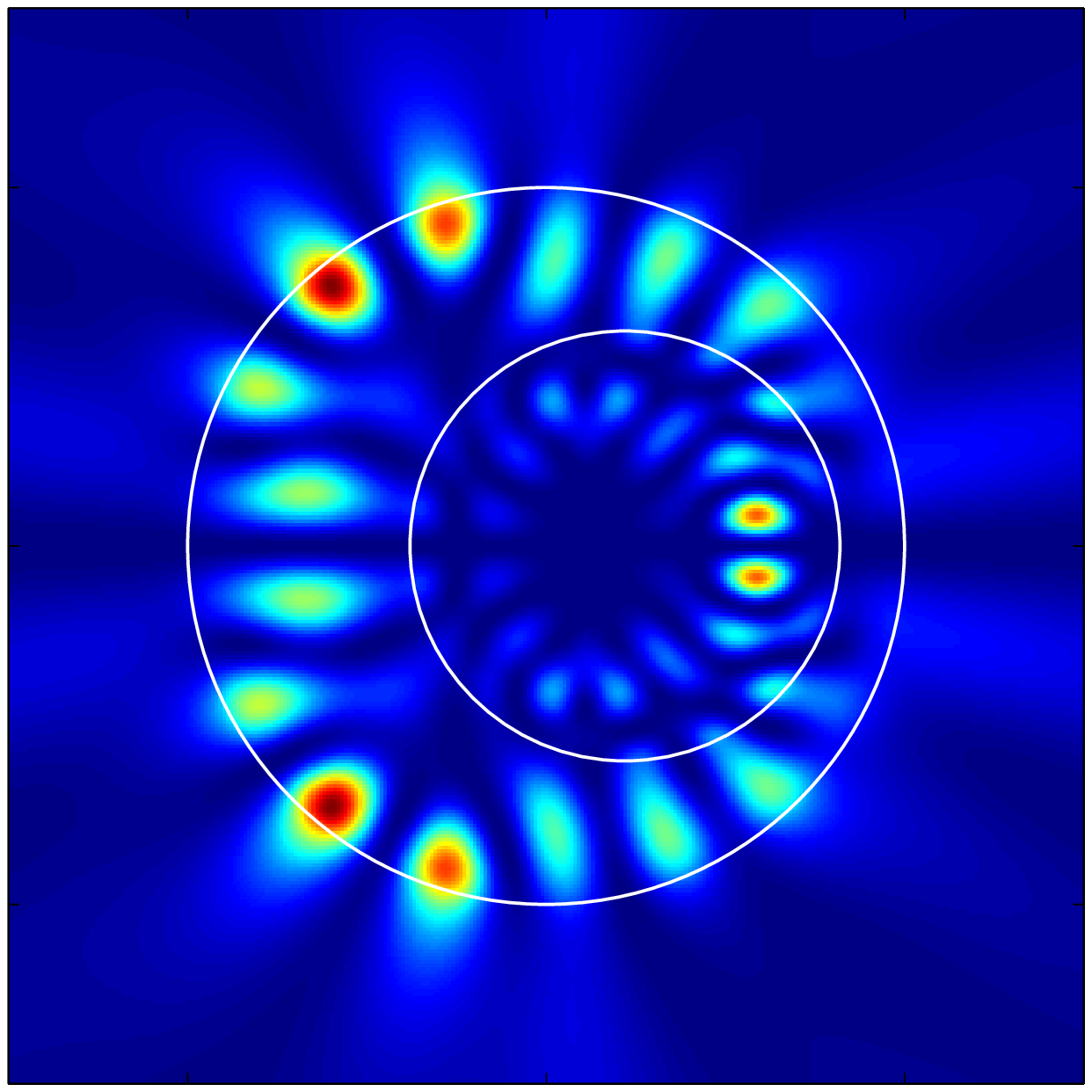}\\[-2.81cm]
\includegraphics[width=8.1cm]{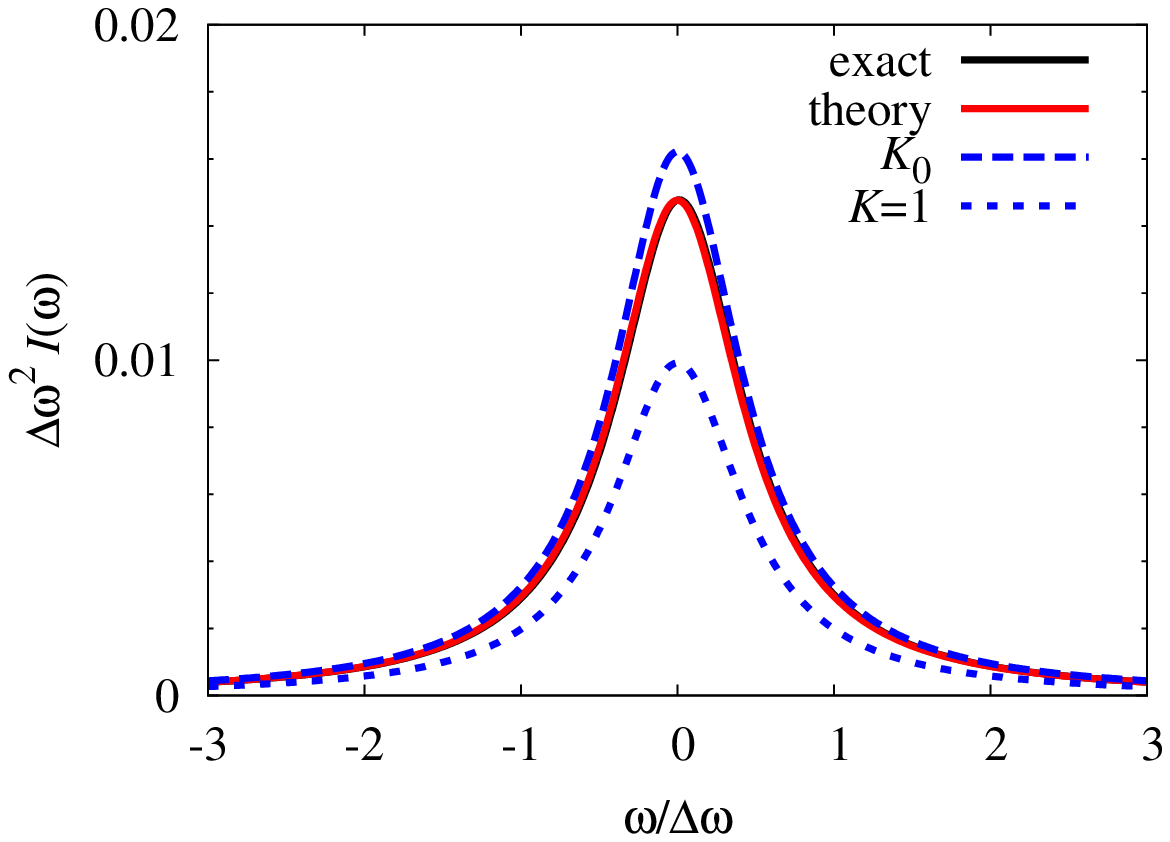}
\caption{(Color) Inset: the white circles indicate the dielectric
interfaces of an annular resonator with radii $R_2=0.6\,R_1$ and
eccentricity $d=0.22\,R_1$. The refractive index in the inner disk
is $n_2=3.3$, while in the annular region $\re n_1$ is fixed to
$1.8$. The color-coded density plot shows a TM-polarized resonator
mode at its threshold ($n_{1,0}=1.8-i0.029$, $\omega_0 R_1=6.45$).
Main panel: frequency-resolved intensity close to the threshold
(suitably scaled to be independent of the small detuning of $\im
n_1$). The exact numerical result (solid black curve) is a
Lorentzian whose width and height can be accurately described
using the theory developed in this paper (solid red curve on top
of the black curve) [Eq.\ (\ref{eq:klorentzian}) with parameters
determined from Eqs.\ (\ref{eq:genkall})-(\ref{eq:coldwc}) (TM
polarization)]. A discrepancy in the height is observed when the
Petermann factor is determined from the homogeneous expression
$K_{\rm 0}$, Eq.\ (\ref{eq:ksp}) (dashed blue curved), or when it
is completely neglected (dotted blue curve, corresponding to the
Schawlow-Townes result).} \label{fig1}
\end{figure}

The past decade has seen the development of a new class of microlasers, in
which the confinement is due to internal reflection at dielectric interfaces
\cite{microresonators1,microresonators2,microresonators3,microresonators4,microresonators5,microresonators6,%
microresonators7,microresonators8}. The typical feature size $s$
is often not much larger than the optical wavelength $\lambda$.
Due to their shape, these resonators support effectively
two-dimensional, highly non-uniform, anisotropic modes which are
of a decisively non-paraxial character. These systems are
currently under intense experimental and theoretical
investigations. Until very recently, however, the consequences of
mode nonorthogonality have been ignored. Models based on
random-matrix theory \cite{patra,frahm,schomerus,keating} only
partially account for the structured wave patterns of typical
microresonators. The first steps into this direction have been
taken in recent numerical investigations of stadium billiards
\cite{kim} and spirals \cite{wiersignew}, which focus on the case
of two cross-talking modes \cite{berry} and demonstrate that the
Petermann factor can be arbitrarily large even for high-$Q$ modes.
Experiments on this subject have just picked up \cite{leeexp}.
Therefore, there is a demand for a systematic theory of mode
nonorthogonality and its consequences for the quantum noise that
accounts for the specific features of dielectric microresonators.

In this paper I fill this gap and develop a theory of the
frequency-resolved light emission that applies to general,
nonparaxial modes in such resonators and furthermore admits
spatial non-uniformity of the dielectric constant and the gain (I
assume single mode lasing and therefore also ignore degeneracies
of lasing modes). The result entails that formally, the concept of
the Petermann factor emerges only in the limit of a small
wavelength (or a large resonator), $\lambda/s\ll 1$. In reality,
however, the relations given below already work for relatively
small resonators (for illustration see Fig.\ \ref{fig1}, which
presents results for an annular microresonator where both the
dielectric constant as well as the gain are nonuniform.)  The
Lorentzian intensity profile can then be parameterized by a few
characteristic parameters,
\begin{equation}
I(\omega)=\frac{K}{2\pi}\frac{\Gamma^2+4\delta\omega_c^2}{(\omega-\omega_0-\delta\omega)^2+\Delta\omega^2/4}.
\label{eq:klorentzian}
\end{equation}
This result contains an appropriately defined Petermann factor in the form of
a generalized condition number,
\begin{subequations}\label{eq:genkall}
\begin{eqnarray}
K&=&\left|\frac{\int|\psi_0|^2\im n_0^2\,d{\bf r}}{\int \psi_0^2\im n_0^2
\,d{\bf r}}
\right|^2 \qquad(\mbox{TM polarization}),\qquad
\label{eq:genk}
\\[.2cm]
K&=&\left|\frac{\int|\nabla\psi_0|^2\im n_0^{-2}\,d{\bf r}}{\int (\nabla\psi_0)^2\im n_0^{-2}
\,d{\bf r}}
\right|^2 \quad(\mbox{TE polarization}),\qquad
\label{eq:genkte}
\end{eqnarray}
\end{subequations}
where $n_0$ is the complex refractive index at threshold, and $\psi_0$ is the
associated resonance wavefunction. The resonance frequency at threshold is
denoted by $\omega_0$. Below threshold, the linewidth is given by
\begin{subequations}\label{eq:pole0}
\begin{eqnarray}
\Delta\omega&=& -2\im\omega_0\frac{\int \psi_0^2n_0 (n_0-n) \,d{\bf r}}{\int
\psi_0^2n_0^2\,d{\bf r}}\quad\mbox{(TM)},\\
\Delta\omega&=& 2\im\frac{\int (\nabla\psi_0)^2n_0^{-3} (n_0-n) \,d{\bf r}}{\omega_0\int
\psi_0^2\,d{\bf r}}\quad\mbox{(TE)},
\end{eqnarray}
\end{subequations}
where $n$ is the refractive index at the working point.
(Throughout this work the vacuum speed of light $c\equiv 1$; above
threshold, the linewidth is further reduced by a factor of 1/2
\cite{goldberg}, as in the Schawlow-Townes case.) The theory also
predicts a systematic line shift
\begin{subequations}\label{eq:dw}
\begin{eqnarray}
\delta\omega&=&\omega_0\re\frac{\int \psi_0^2n_0 (n_0-n) \,d{\bf r}}{\int
\psi_0^2n_0^2\,d{\bf r}}\quad\mbox{(TM)},\\
\delta\omega&=& -\re\frac{\int (\nabla\psi_0)^2n_0^{-3} (n_0-n) \,d{\bf r}}{\omega_0\int
\psi_0^2\,d{\bf r}}\quad\mbox{(TE)},
\end{eqnarray}
\end{subequations}
which is of the same order as the linewidth and generally only disappears for
large resonators filled with a homogeneous medium. The cold-cavity
characteristics $\Gamma$ and $\delta\omega_c$ are obtained from the relation
$\delta\omega_c-i\Gamma/2=\delta \Omega_c$ where
\begin{subequations}
\label{eq:coldwc}
\begin{eqnarray}
\delta\Omega_c&=&i\frac{\omega_0\int \psi_0^2\im n_0^2\,d{\bf r}}{2\int
\psi_0^2 n_0^2\,d{\bf r}}\quad\mbox{(TM)},
\\
\delta\Omega_c&=&-i\frac{\int (\nabla\psi_0)^2\im n_0^{-2}\,d{\bf
r}}{2\omega_0\int\psi_0^2\,d{\bf r}}\quad\mbox{(TE)}.
\end{eqnarray}
\end{subequations}
As a notable side product, the latter expressions generalize and improve the
common phenomenological relation $\im n_0=-(2  \Gamma/\omega_0)\re n_0$ (TM)
between the cold-cavity lifetime and the laser threshold of homogenous
systems.

All considerations in this paper are based on the effective-medium
approach, where the geometric and material-specific properties of
a dielectric microresonator are described by a position-dependent
refractive index $n({\bf r})$. Let us assume $n({\bf r})=1$
outside the resonator. Inside the resonator the refractive index
can be complex, with $\im n<0$ in the amplifying (active) regions
of the medium. For an effectively two-dimensional resonator, the
classical electromagnetic field is represented by a scalar
wavefunction $\psi({\bf r};\omega)$ which fulfills the Helmholtz
equation \cite{vassallo}
\begin{subequations}\label{eq:helm}
\begin{eqnarray}
{\cal L} \psi({\bf r};\omega)=0,&&{\cal
L}=\Delta+
\omega^2
n^2({\bf r}) \qquad\mbox{(TM)},
\label{eq:helmtm}
\\
&& {\cal L}=\nabla n^{-2}({\bf r}) \nabla+\omega^2 \quad\mbox{(TE)}.
\label{eq:helmte}
\end{eqnarray}
\end{subequations}
Consider the case that $\re n({\bf r})$ is fixed while $\im n({\bf
r})$ can be controlled via pumping. At the laser threshold $\im
n({\bf r})\equiv\im n_0({\bf r})$, Eq.\ (\ref{eq:helm}) admits a
solution $\psi_0$ satisfying purely outgoing boundary conditions
with real angular frequency $\omega_0$. Below the laser threshold,
the resonance frequency $\Omega$ associated to this solution
becomes complex, where $-2\im \Omega=\gamma$ is the decay rate.
The cold-cavity decay rate $\Gamma$ is obtained for $\im n=0$ (the
complex cold-cavity resonance frequency is denoted by
$\Omega_c=\omega_c-i\Gamma/2$).

Quantum optics adds spontaneously emitted photons to this picture, which are
generated and amplified by the active medium. Even in absence of external
illumination, the resonator then emits photons of frequency $\omega\approx
\re\Omega$. A convenient quantum-optical framework to study this radiation is
provided by the input-output formalism, which delivers the output intensity
\cite{beenakker,schomerus}
\begin{equation}
I(\omega)=\frac{1}{2\pi}\tr(S^\dagger S-\openone)
\label{eq:iss}
\end{equation}
in terms of the scattering matrix $S(\omega)$ of the classical field
(assuming complete population inversion in the medium).

For the cold cavity the scattering matrix is unitary, and the
passive resonator does not emit any radiation. For a finite gain
the scattering matrix departs from unitarity
 because the particle flux is no longer conserved due to the
stimulated emission of photons in the active medium. Even if the
cold-cavity resonances are strongly overlapping for the cold
cavity (as is typical for low-$Q$ modes with $\Gamma$ larger than
the mode spacing), well-resolved resonances appear when one steers
the system close to the laser threshold. Dielectric microresonator
design aims to support high-$Q$ modes, which can be resolved even
deep below threshold \cite{kim,leeexp}; in that case, however, the
threshold is reached more quickly than in the case of low-$Q$
modes. Both cases can therefore be approached by a systematic
expansion around the threshold condition \cite{epaps1}, which
delivers the output intensity

{\small
\begin{subequations}\label{eq:iout}
\begin{eqnarray}
&&I(\omega)=\frac{1}{2\pi}\left|\frac{\int |\psi_0|^2\im
n_0^2\,d{\bf r}}{\int
\psi_0^2n_0^2(\omega/\omega_0+n/n_0-2)\,d{\bf r}}\right|^2\quad\mbox{(TM)},
\label{eq:iouttm}\qquad
\\
&&I(\omega)=\frac{1}{2\pi}\left|\frac{\int |\nabla\psi_0|^2\im
n_0^{-2}\,d{\bf r}} {
\omega_0(\omega-\omega_0)\int\psi_0^2\,d{\bf r}+
\int (\nabla\psi_0)^2\frac{n-n_0}{n_0^3}\,d{\bf r}}
\right|^2\,\,\mbox{(TE)}.
\nonumber\\
\label{eq:ioutte}
\end{eqnarray}
\end{subequations}
} This expression formally diverges at the complex frequency
$\Omega=\omega_0+\delta\Omega$ where
\begin{subequations}\label{eq:pole}
\begin{eqnarray}
\delta\Omega&=&\omega_0\frac{\int \psi_0^2n_0 (n_0-n) \,d{\bf r}}{\int
\psi_0^2n_0^2\,d{\bf r}}\qquad\mbox{(TM)},\\
\delta\Omega&=& -\frac{\int (\nabla\psi_0)^2n_0^{-3} (n_0-n) \,d{\bf r}}{\omega_0\int
\psi_0^2\,d{\bf r}}\quad\mbox{(TE)},
\end{eqnarray}
\end{subequations}
The output intensity is hence a Lorentzian [Eq.\
(\ref{eq:lorentzian})] whose center $\bar\omega=\re\Omega$ is
displaced from $\omega_0$ by a systematic line shift $
\delta\omega=\re \delta \Omega$ [see Eq.\ (\ref{eq:dw})],
 which is generally of the same order as
the width $\Delta\omega=-2\im \delta \Omega$ of the Lorentzian [see Eq.\
(\ref{eq:pole0})]. The latter can be related to the total intensity $I=\int
I(\omega)\,d\omega$ via the relation
\begin{subequations}\label{eq:dogen}
\begin{eqnarray}
\Delta\omega&=&\frac{1}{I}\left|\frac{\omega_0\int|\psi_0|^2\im
n_0^2\,d{\bf r}}{\int \psi_0^2n_0^2 \,d{\bf r}}
\right|^2\quad\mbox{(TM)},\\
\Delta\omega&=&\frac{1}{I}\left|\frac{\int|\nabla\psi_0|^2\im
n_0^{-2}\,d{\bf r}}{\omega_0\int \psi_0^2 \,d{\bf r}
}\right|^2\quad\mbox{(TE)}.
\end{eqnarray}
\end{subequations}
This expression contains integrals similar to those in the
expression for $K_{\rm 0}$ [Eq.\ (\ref{eq:ksp})],  but weighted by
the refractive index. Moreover, instead of the transverse mode
profile, $\psi_0$ now represents the resonance wavefunction in the
resonator plane, which encodes the full, non-paraxial mode
structure. At this point, the integral in the denominator is not
restricted to the interior of the resonator, and the result does
not feature the cold-cavity decay rate. For the latter reason Eq.\
(\ref{eq:dogen}) cannot be used to extract a generalized Petermann
factor.

In order to make contact with the conventional theory of the
linewidth we now consider the regime
 $\lambda\ll s$ of a large
resonator. Under this condition, typical resonances have a small
cold-cavity decay rate, $\Gamma\ll \omega_0$, and the laser
threshold is attained at a small gain, $|\im n_0|\ll \re n_0$.
Equation (\ref{eq:pole}) then applies all the way down to the
cold-cavity limit, and can be further recast to deliver the
complex cold-cavity resonance frequency
$\Omega_c=\omega_0+\delta\Omega_c$, where $\delta\Omega_c$ is
given by Eq.\ (\ref{eq:coldwc}). This result is perturbative in
the gain but nonperturbative in the openness of the system, which
is encoded in $\omega_0$, $n_0$, and $\psi_0$. Note that in  Eq.\
(\ref{eq:coldwc}), the numerator is determined by the active
region, while the denominator also depends on the passive regions
--- including the region outside the resonator, where integrals are well
defined because $n$ is real \cite{regularization}. For homogeneous systems,
the simple relation $\Omega_c=\omega_0+i(\im n_0/\re n_0)\omega_0$ (TM) is
often applied (this also entails the relation for the laser threshold
discussed above). In comparison to Eq.\ (\ref{eq:coldwc}), this simple
expression is only valid when one can ignore to the exterior contribution to
the denominator, which in practice requires a very large system size.

With Eq.\ (\ref{eq:coldwc}),  linewidth (\ref{eq:dogen})
simplifies to
\begin{subequations}
 \label{eq:finalexpression}
\begin{eqnarray}
\Delta\omega&=&\frac{\Gamma^2+4\delta\omega_c^2}{I}\left|\frac{\int|\psi_0|^2\im
n_0^2\,d{\bf r}}{\int \psi_0^2\im n_0^2 \,d{\bf r}
}\right|^2\quad\mbox{(TM)},\\
\Delta\omega&=&\frac{\Gamma^2+4\delta\omega_c^2}{I}\left|\frac{\int|\nabla\psi_0|^2\im
n_0^{-2}\,d{\bf r}}{\int (\nabla\psi_0)^2\im
n_0^{-2} \,d{\bf r}
}\right|^2\quad\mbox{(TE)},\quad
\end{eqnarray}
\end{subequations}
which features the cold-cavity decay rate
$\Gamma=-2\im\delta\Omega_c$ along with the cold-cavity line shift
$\delta\omega_c=\re\delta\Omega_c$. (Far above the laser
threshold, the linewidth is again halved because amplitude
fluctuations are suppressed by the nonlinear feedback with the
medium \cite{goldberg}). The Petermann factor, hence, takes the
form of Eq.\ (\ref{eq:genkall}). For TM polarization, the
conventional form [Eq.\ (\ref{eq:ksp})] of the Petermann factor is
recovered when the active medium fills the resonator homogeneously
(so that $n_0$ does not depend on position). This conceptual
relation hence applies even when the mode structure in the
resonator plane is strongly non-paraxial.  This is also the case
for TE polarization or an inhomogeneous medium, where Eq.\
(\ref{eq:genkall}) relates the excess noise to an appropriately
generalized measure of mode nonorthogonality (satisfying $K\to1$
in the limit of a closed resonator). For an inhomogeneous medium,
Eq.\ (\ref{eq:genkall}) shows that the excess noise is generated
in the active regions (where $\im n_0\neq 0)$.

{\em Model system}.{} In order to assess the predictive power of
the above analysis I turn to a paradigmatic example of a
microresonator with a nonuniform refractive index, which at the
same time displays a rich variety of resonator modes. This is
provided by the annular microresonator geometry, defined by two
eccentric circular interfaces such as indicated in the inset of
Fig.\ \ref{fig1}. In order to exploit the full generality of the
formalism presented above, I assume that the gain is constrained
to the annular region between these interfaces ($\im n_1\equiv
-i\tilde n$), but vanishes in the interior disk ($\im n_2=0$), and
furthermore set $\re n_1=1.8$ and $\re n_2=3.3$. The radii of the
two circles are related by $R_2=0.6\,R_1$, and the eccentricity is
$d=0.22\, R_1$. The finite eccentricity breaks the rotational
symmetry and hence leads to nonintegrable classical ray dynamics.
The classical phase space accommodates domains of stability
embedded into regions of chaotic instability and, hence, exhibits
the full complexity of generic dynamical systems. The
corresponding wave dynamics reflect this complexity in a rich set
of resonance wavefunctions, which also includes examples with a
highly directional far-field radiation pattern
\cite{hentschel,wiersig}.

Numerical techniques make it possible to investigate a large
number of modes, especially in the interesting regime where
$\omega_0R_1$ is moderately large \cite{hentschel}. Here I focus
on a representative example of a resonator mode with TM
polarization,  which reaches its threshold at $\tilde
n_{0}=0.029$, $\omega_0 R_1=6.45$. In the main panel of Fig.\
\ref{fig1}, the solid black curve shows the intensity profile
close to threshold (for a small detuning of $\tilde n$), which is
directly obtained from Eq.\ (\ref{eq:iss}) via a numerical
computation of the scattering matrix. The curve is scaled so that
it becomes independent of the detuning when $\tilde n-\tilde
n_{0}$ is small. The red curve is obtained from the theory
developed in this work: a Lorentzian of   form
(\ref{eq:klorentzian}), where the parameters are determined by
Eqs.\ (\ref{eq:genkall})-(\ref{eq:coldwc}) (for TM polarization).
This curve lies on top of the black curve so that the latter is
barely visible. The theory of this paper hence applies even though
the resonator is only moderately larger than the wavelength. The
other curves illustrate the necessity to adopt the results
developed in this paper. The dashed blue curve is obtained when
the Petermann factor is taken of the homogeneous form $K_{\rm 0}$
[Eq.\ (\ref{eq:ksp})], while the dotted blue curve shows the
result when mode nonorthogonality is entirely ignored (as in
Breit-Wigner theory, which delivers the Shawlow-Townes formula)
\cite{epaps2}.

For completeness I remark that in the case of a homogeneous
circular disk resonator (corresponding to $R_2= 0$), Eq.\
(\ref{eq:ksp}) applies. In polar coordinates, the radial
dependence is given by Bessel functions, while the constraint on
time-reversal symmetry
 singles out modes with a standing-wave angular dependence
 $\sin(m(\phi-\phi_0))$ \cite{hentschel}.
For TM polarization and $\omega R \gg 1$,  Petermann factor
(\ref{eq:ksp}) then takes  the asymptotic form $K=\sinh^2x/x^2$
where $x=2\im (n \omega R_1)$ and therefore approaches unity for
modes with a large $Q$ factor.

In summary, I have explored the consequences of mode
nonorthogonality in dielectric microresonators for the general
case of a highly non-paraxial mode structure and nonuniform
material properties. This theory leads to an appropriately
generalized Petermann factor [Eq.\ (\ref{eq:genkall})]. Compact
expressions (\ref{eq:klorentzian})-(\ref{eq:coldwc}) can be used
to inform the interpretation of experiments beyond the simple
effective phenomenological models employed to date.
 An open question (beyond the illustrative example of Fig.\
\ref{fig1}) that can be pursued on basis of these expressions is
how complex quantum dynamics (regular versus chaotic wave
patterns) express themselves in the mode nonorthogonality.

This work was supported by the European Commission via
Marie-Curie Excellence Grant No.\ MEXT-CT-2005-023778
(Nanoelectrophotonics).


\pagebreak
\appendix
\section{Further details of the derivation\label{appa}} This appendix details some technical steps
in the derivation of Eq.\ (\ref{eq:iout}). The analysis exploits
that near resonance, the outgoing radiation is dominated by the
resonance wavefunctions, $\psi\approx \psi_0$. To be specific, let
us consider a disk $\cal C$ of radius $R\to\infty$ that completely
covers the resonator, so that outside the disk $n({\bf r})=1$.
Using polar coordinates $r$, $\varphi$ in a coordinate system with
origin at the center of the disk, a basis of scattering
wavefunctions is then given by
\begin{equation}\label{eq:basis}
\phi_m^{(\sigma)}(r,\varphi;\omega)= H_{m}^{(\sigma)}(\omega
r)e^{i\sigma m\varphi},
\end{equation}
where $H$ is the Hankel function and $m=0,\pm1,\pm2,\ldots$. The index
$\sigma$ takes the value $\sigma=+$ for outgoing scattering states while
$\sigma=-$ for incoming scattering states.

The expansion of the scattering wavefunction in the basis (\ref{eq:basis})
defines, respectively, the in- and outgoing scattering amplitudes  $a_m$ and
$b_m$,
\begin{eqnarray}\label{eq:basis2}
\psi({\bf r})&=&\sum_m\left[ b_m
\phi_m^{(+)}(r,\varphi)
+a_m\phi_m^{(-)}(r,\varphi)\right].
\end{eqnarray}
Adopting a vector notation the amplitudes are related via the scattering
matrix $S$, ${\bf b}=S {\bf a}$, which is found by matching the scattering
states to solutions of Eq.\ (\ref{eq:helmtm}) inside the disk. The expansion
coefficients of the resonance wavefunction in the scattering region are
denoted by ${\bf b}_0$.

The following considerations rest on a consequence of the divergence theorem,
\begin{equation}\label{eq:div}
\int_{\cal C} d{\bf r}\, \psi_1\Delta\psi_2=\int_{\cal C} d{\bf
r}\, \psi_2\Delta\psi_1+\oint_{\partial \cal C}d\varphi\, R
\left[\psi_1\frac{\partial \psi_2}{\partial r}-\psi_2\frac{\partial
\psi_1}{\partial r}\right],
\end{equation}
which holds for arbitrary well-behaved functions $\psi_1$, $\psi_2$, and
 the orthogonality property
\begin{eqnarray}\label{eq:orth}
&&\oint_{\partial \cal C}d\varphi\, R \left[\phi_m^{(\sigma)}  \frac{\partial
\phi_{m'}^{(\sigma')}}{\partial r}  -\phi_{m'}^{(\sigma')} \frac{\partial
\phi_m^{(\sigma)}}{\partial r} \right]
=4i(\sigma'-\sigma)\delta_{ m,m'}  \nonumber\\&&
\end{eqnarray}
of the scattering basis functions ($\partial C$ denotes the circular boundary
of the disk ${\cal C}$, which eventually moves to infinity).

Due to time-reversal symmetry, the scattering matrix is symmetric, $S=S^T$.
To see that this symmetry holds for complex $n$, consider two arbitrary
solutions $\psi_1$, $\psi_2$ of the Helmholtz equation, and apply the
definition of the scattering matrix as well as Eqs.\ (\ref{eq:div}) and
(\ref{eq:orth}) to evaluate
\begin{equation}
-8i{\bf a}_2^T(S-S^T){\bf a}_1=\int_{\cal C}(\psi_2{\cal L}\psi_1-\psi_1{\cal L}\psi_2)\,d{\bf r}
=0.
\end{equation}

Close the resonance, the outgoing radiation will be dominated by the
characteristics of the resonant state. Hence, $S\approx\alpha {\bf b}_0{\bf
b}_0^T$, where $\alpha$ diverges at the resonance so that other outgoing wave
components can be ignored.

The evaluation of the coefficient $\alpha$ requires to compute the overlap
between the resonance wavefunction $\psi_0$ and an arbitrary near-resonant
solution $\psi\approx \psi_0$, which also possesses a small incoming
component. Assume that $\psi$ is normalized so that the dominant outgoing
radiation is described by the amplitudes ${\bf b}_0$. The scattering matrix
then takes the form $S\approx {\bf b}_0{\bf b}_0^T/ ({\bf b}_0^T{\bf a})$,
while the intensity becomes
\begin{equation}
I\approx\frac{1}{2\pi}\frac{|{\bf b}_0^\dagger{\bf b}_0|^2}{|{\bf b}_0^T{\bf a}|^2}.
\label{eq:ires}
\end{equation}
In both expressions the denominator is independent of the choice of the
near-resonant wavefunction $\psi$:
\begin{eqnarray}\label{eq:id1}
-8i{\bf b}_0^T{\bf a}&=&\int[\psi_0(\Delta\psi)-\psi(\Delta\psi_0)]\,d{\bf r}
\nonumber\\ &\approx&\int\psi_0^2(n^2\omega^2-n_0^2\omega_0^2)\,d{\bf r},
\end{eqnarray}
where in the last step we used the Helmholtz equation. As we are close to
resonance, $n^2\omega^2-n_0^2\omega_0^2\approx 2
n_0^2\omega_0^2(\omega/\omega_0+n/n_0-2)$.

We now once more exploit time-reversal symmetry, this time in the form that
$\psi_0^*$ provides a purely incoming solution of the Helmholtz equation
where $n_0\to n_0^*$. Therefore,
 \begin{eqnarray} \label{eq:id2}
-8i{\bf b}_0^\dagger{\bf b}_0&=&\int[\psi_0(\Delta\psi_0^*)-\psi_0^*(\Delta\psi_0)]\,d{\bf r}
\nonumber \\
&=&\int|\psi_0|^2(n_0^2-{n_0^*}^2)\omega^2_0\,d{\bf r}.
\end{eqnarray}

Equation (\ref{eq:iouttm}) now follows from Eqs.\ (\ref{eq:ires}),
(\ref{eq:id1}), and (\ref{eq:id2}). The techniques described above
can also be adapted to modes with TE polarization which are
governed by the Helmholtz equation (\ref{eq:helmte}); this then
results in  Eq.\ (\ref{eq:ioutte}).

\newpage \begin{widetext}
\section{Comparison to other resonator modes\label{appb}}

\begin{figure*}[b]
\includegraphics[width=\textwidth]{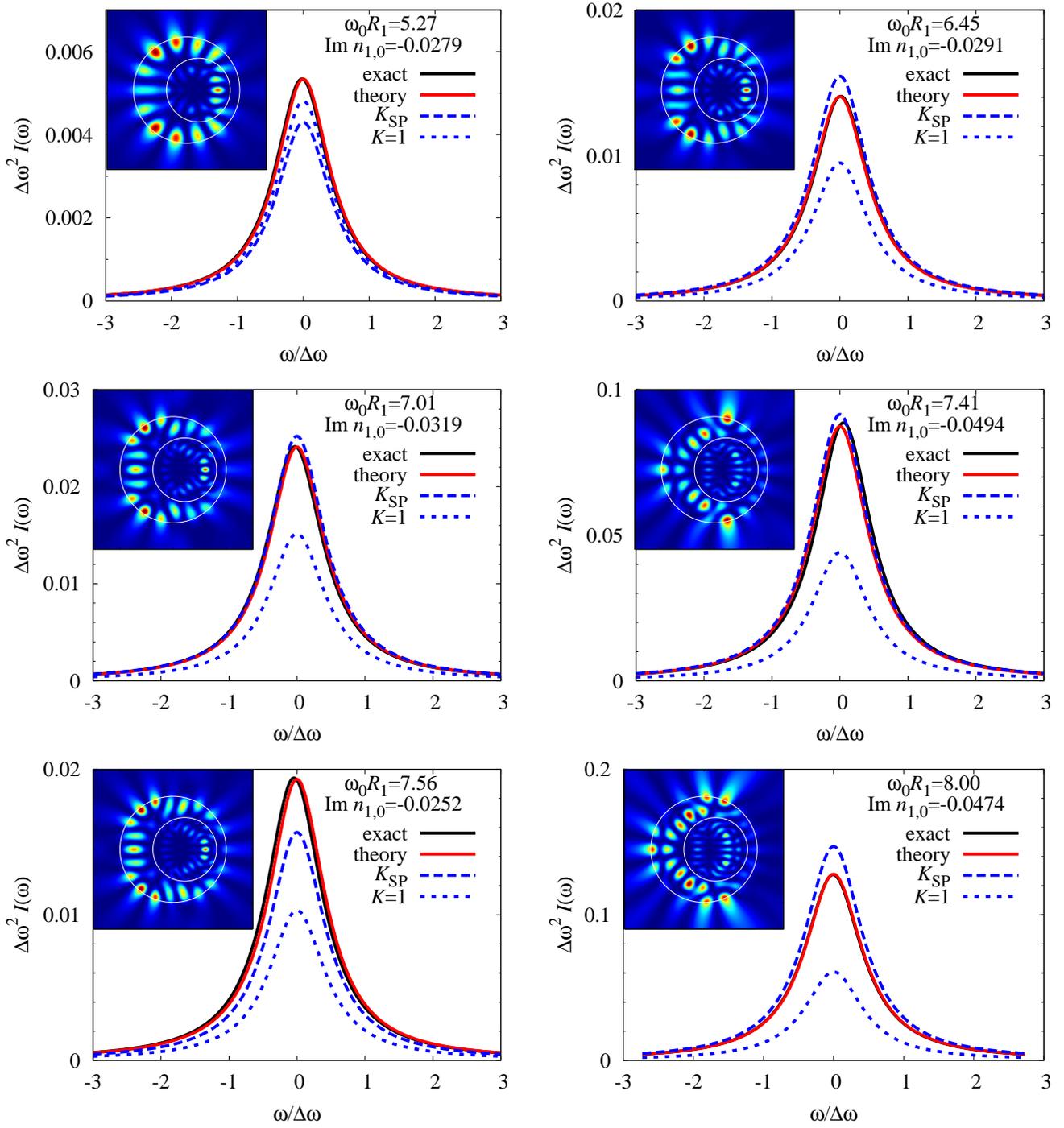}
\caption{(Color) Same as Fig.\ \ref{fig1}, but for other resonator
modes of the annular resonator. All modes have even parity with
respect to the horizontal resonator axis. The
 mode with $\omega_0 R_1=6.45$ is almost
degenerate with the odd-parity mode of Fig.\ \ref{fig1}.}
\label{fig2}
\end{figure*}
\pagebreak
\end{widetext}
\end{document}